\documentclass[journal=jctcce,manuscript=article]{achemso}
\usepackage{amsmath}
\usepackage{amssymb}
\usepackage{graphicx}
\usepackage{color}
\usepackage[normalem]{ulem}
\usepackage[table]{xcolor}

\date{\today}

\newcommand{\gt}[1]{{ #1 }}

\newcommand{\avol}[1]{{ #1 }}

\newlength{\imgsz}
\title{Probing the unfolded configurations of a beta hairpin using sketch-map}

\author{Albert Ardevol}
\email{albert.ardevol@phys.chem.ethz.ch}
\affiliation
{Computational Science, Department of Chemistry and Applied Biosciences,
ETH Zurich, USI-Campus,
Via Giuseppe Buffi 13, C-6900 Lugano, Switzerland}
\author{Gareth A. Tribello}
\affiliation
{Atomistic Simulation Centre, School of Mathematics and Physics, 
Queen's University Belfast, Belfast, BT7 1NN, United Kingdom}

\author{Michele Ceriotti}
\affiliation
{Laboratory of Computational Science and Modelling, 
EPFL, Switzerland}

\author{Michele Parrinello}
\affiliation
{Computational Science, Department of Chemistry and Applied Biosciences, 
ETH Zurich, USI-Campus, 
Via Giuseppe Buffi 13, C-6900 Lugano, Switzerland}

\begin{document}

\singlespacing
\begin{abstract}
This work examines the conformational ensemble involved in $\beta$-hairpin folding 
by means of advanced molecular dynamics simulations and dimensionality reduction.  A fully 
atomistic description of the protein and the surrounding solvent molecules is used and this complex 
energy landscape is sampled by means of parallel tempering metadynamics 
simulations.  The 
ensemble of configurations explored is analysed using the recently proposed sketch-map algorithm.  
Further simulations allow us to probe how mutations affect the structures adopted by this protein.
We find that many of the configurations adopted by a mutant are the 
same as those adopted by the wild type protein.  Furthermore, certain mutations 
destabilize secondary structure containing configurations by preventing the 
formation of hydrogen bonds or by promoting the formation of new intramolecular 
contacts. Our analysis demonstrates that machine-learning 
techniques can be used to study the energy landscapes of complex molecules and 
that the visualizations that are generated in this way provide a natural basis 
for examining how the stabilities of particular configurations of the molecule 
are affected by factors such as temperature or structural mutations.
\end{abstract}

\section{Introduction}

Protein molecules are the workhorses of the cell and are responsible for biological functions ranging 
from enzymatic catalysis to cell motility \cite{bio-book}.  In many proteins this functionality is connected to the fact 
that the protein adopts a very specific tertiary structure and in much of
biochemistry a protein's 
function and behavior in the cell is rationalized by referring to specific
details in the protein's 
static, tertiary structure.  Although this approach has been very successful, there is a growing 
consensus that this static picture of protein function is often incomplete.  Processes such as signalling \cite{signal} 
and allosteric binding \cite{allosteric} as well as the the growing class of so-called intrinsically disordered proteins \cite{idp}
all suggest that the dynamical behaviour of proteins is important and that this must be incorporated 
when considering protein function.  

This requirement to understand both the average (static) structure of the protein and its dynamical 
behavior presents particular problems to the experimentalist. The first concerns how to 
extract time-resolved structural information, as opposed to time-averaged information, from experiments.  
Simulations help enormously in this regard by providing tools that allow one to ``watch" protein motions 
in real time \cite{long-md-simulation}.  That said, explicit all-atom simulations of proteins provide almost too much information.  
In a typical molecular dynamics (MD) simulation the changes in the positions of 
all the atoms in the protein 
as a function of time is calculated.  The trajectory that is output when a 
system of $N$ atoms is simulated thus consists of an ordered set of 
$3N$-dimensional vectors. The problem then, when analysing the trajectory, is 
that the 
information on the interesting long-timescale events (e.g. protein folding or conformational changes) is 
hidden a sea of coordinate information on what are for the most 
part uninteresting, short-time scale events (e.g. bond vibrations).    
What we would really like to do is to obtain a representation of the trajectory 
that is based on a small number of variables, which differentiate between 
structures that inter-convert on the timescale of interest. Any remaining 
variables are those that differentiate between structures that inter-convert 
more 
rapidly - these can be safely integrated out.        

A second difficulty that arises with theories that incorporate the dynamical behaviour of the protein as 
well as its static structure concern the explanation of mutagenesis data.  When a protein's mode of 
operation can be rationalized based only on the averaged tertiary structure it is easy to visualize, 
and hence understand, how mutations affect structure and hence functionality \cite{mutagenesis}. By contrast, if the 
dynamical behavior is important, it becomes far more difficult to explain why the functionality changes 
upon mutation.  The average structures of the mutant and wild type could now be 
the same and any differences 
in functionality might only arise because the mutant subtly perturbs the energy landscape and makes it more/less 
favourable for the protein to adopt some special, higher-energy, biological 
active form \cite{trans-lyzo}.  Atomistic simulations 
can again play a role when it comes to phenomena like this.  In addition, the question to be addressed in such 
simulations is articulated more clearly.  \gt{We simply have to find whether 
there are particular configurations of the protein that have free energies that 
are comparable with the free energy of the folded state for the wild type and 
how the free energies of these structures relative to the folded state is 
affected by the mutation.}

In this paper we show one way that simulation can address this question of how a mutation affects protein function.  
The essence of our approach is to use long parallel tempering metadynamics 
\cite{PTmetad} simulations to explore
{the part of configuration space that is energetically-accessible to both
 the wild type and the mutant. \avol{Parallel tempering methods 
have been extensively used in enhanced sampling simulations of small peptides\cite{gnanakaran, weinstock, deighan}
because they tend to have a relatively low melting temperature and a fast conformational 
diffusion in the unfolded ensemble, which are good characteristics for efficient 
performance of parallel tempering.\cite{zuckerman}} The output from these simulations is 
high-dimensional and difficult to interpret, which we resolve by using the sketch-map algorithm \cite{sketch-map} to generate a 
two-dimensional map of configuration space.  Free energy surfaces as a function of these coordinates give insight 
into the likelihood of the protein adopting a particular configuration. As such, differences in the behavior of the 
mutant and wild type sequences can be understood by comparing the free energy surfaces obtained from the simulations 
of the wild type and the mutant.

\section{Background}

There are a number of non-linear dimensionality reduction (NLDR) algorithms 
\cite{isomap,lle,diffmap-1,diffmap-2,diffmap-3} that are now used almost 
routinely to generate 
low-dimensionality representations of more high-dimensional information. In 
these algorithms a computer is used to fit the parameters of some non-linear 
function, which transforms the high-dimensional coordinates to a 
lower-dimensional vector. When using  \gt{any dimensionality 
reduction} algorithm, and in particular when 
choosing the particular non-linear function to fit, one is forced to make 
assumptions about the structure of the high-dimensional data. As a 
consequence \gt{some of} these algorithms \gt{may}  not  
necessarily \gt{be} well-suited  for treating the sort of data one
extracts from a typical molecular dynamics (MD) or enhanced sampling 
trajectory. The sketch-map dimensionality reduction algorithm
\cite{sketch-map,fieldcvs} was recently developed with the
 features of the configurational landscape of 
atomic and molecular systems that perhaps make using these other algorithms 
problematic in mind.  In particular, sketch-map
disregards the information that corresponds to 
thermal fluctuations around a (meta)stable structure. 
This is achieved by doing a form of multidimensional
scaling (MDS) \cite{mds} in which projections for a 
set of $N$ high-dimensionality, landmark points are found by 
minimizing the following stress function:
\begin{equation}
\begin{aligned}
 \chi^2 & = \sum_{i \ne j} \left[ F(R_{ij}) - f(r_{ij}) \right]^2 \\
 \textrm{where} \qquad f(r) & = 1 - ( 1 + 2^{a/b} - 1)(r/\sigma)^a
)^{-b/a} \\
\textrm{and} \qquad F(R) & = 1 - ( 1 + 2^{A/B} - 1)(R/\sigma)^A
)^{-B/A} 
\end{aligned}
\label{eqn:smap}
\end{equation}
where $R_{ij}$ is the dissimilarity between landmark points $\mathbf{X}_i$
and $\mathbf{X}_j$ and $r_{ij}$ is the distance between their projections
$\mathbf{x}_i$ and $\mathbf{x}_j$.  In this work we have measured dissimilarities between 
protein configurations by measuring how much the full set of protein backbone dihedral
angles change on moving between the two configurations.  \sout{We then tuned the 
value of $\sigma$ in the
sigmoid functions, $F(R)$ and $f(r)$, so that the algorithm focuses on
reproducing the distances between points that are in basins that appear to be connected
by a single transition state.} \gt{We tune the values of the parameters in 
the sigmoid functions, $F(R)$ and $f(r)$, using the methods described in the 
appendices of our previous work \cite{cluster-smap}.}  \sout{We did this 
because} The form of these sigmoid
functions ensures that close together points, that are most likely in the same
basin, are then projected close together while points that are far apart, and are
thus likely to be in basins that are not connected by a single transition state,
are projected far apart. Furthermore, by selecting different $a$ parameters for
the high-dimensional and low-dimensional sigmoid functions, $F(D_{ij})$ and
$f(d_{ij})$ respectively, one can alleviate the problems that arise when
attempts are made to project the high-dimensionality features that are present
in the basins in the low-dimensionality space \cite{cluster-smap}.  

Histograms, and free energy surfaces, can be constructed as a function of
sketch-map coordinates because, once projections of the initial landmark 
frames have been determined, the projection, $\mathbf{x}$ of any point in the 
high-dimensionality
space, $\mathbf{X}$, can be found by minimizing:
\begin{equation}
\delta^2(\mathbf{x}) = \sum_{i=1}^N \left\{ F[R_i(\mathbf{X})] - f[
r_i(\mathbf{x}) ] \right\}^2
\label{eqn:out-of-sample}
\end{equation}  
where $R_i(\mathbf{X})$ is the dissimilarity between $\mathbf{X}$ and the $i$th
landmark point and $r_i(\mathbf{x})$ is the distance between $\mathbf{x}$ and
the projection of the $i$th landmark point. In a recent paper
\cite{cluster-smap} we showed that
this procedure is remarkably robust.  In particular, we found that we could 
use sketch-map coordinates constructed using $N$ landmark frames to project 
configurations that were markedly different from all of the landmarks.  
Furthermore, if the two configurations being projected were also markedly 
different from each other they would be projected at two well-separated 
locations in the sketch-map plane.  In other words, sketch-map coordinates can 
differentiate between distinct structures even if these configurations are 
not represented in the set of landmarks.  This feature is crucial in this work, 
as it gives us the confidence that coordinates constructed for the wild type 
can be used to understand the configurations adopted by a mutant.  This mutant
may well adopt configurations that are energetically inaccessible to the wild 
type and that are thus not represented in the set of landmark frames.

\section{Results}

\subsection{Simulations of protein wild type}

In this study we have examined the 16-residue C-terminal fragment of the
immunoglobulin binding domain of B1 of protein G of Streptococcus protein in
explicit solvent (amino acids sequence Ace-GEWTYDDATKTFTVTE-NMe) \cite{hairpin1,hairpin2,hairpin3}. We used
gromacs-4.5.5 \cite{gromacs} and the variant on replica exchange discussed by Deighan \emph{et
al.} \cite{deighan}.  In this protocol a 100~ns parallel tempering well-tempered ensemble
metadynamics \cite{metad} simulation was run with 32 replicas which had 
temperatures
distributed between between 268 K and 625 K.  In this first set of simulations a
history dependent bias is added that is a function of the potential energy of
the system.  Biases as a function of this collective variable have been shown to
enhance the fluctuations in the energy \cite{wte}.  As such using this form of biasing in
tandem with parallel tempering allows one to lower the number of replicas
required, while ensuring that the exchange probabilities remain reasonable. Once
these simulations were completed we ran 300~ns/replica long well-tempered
metadynamics parallel tempering simulations. In these calculations the bias on
the potential was kept constant and further metadynamics
biases were added on the radius of gyration and the number of hydrogen bonds
between backbone atoms.
  All metadynamics calculations were run using PLUMED 1.3 \cite{plumed,plumed2}
and input files are provided in the supporting information.

\begin{figure}
\centering
 \includegraphics[width=0.9\textwidth]{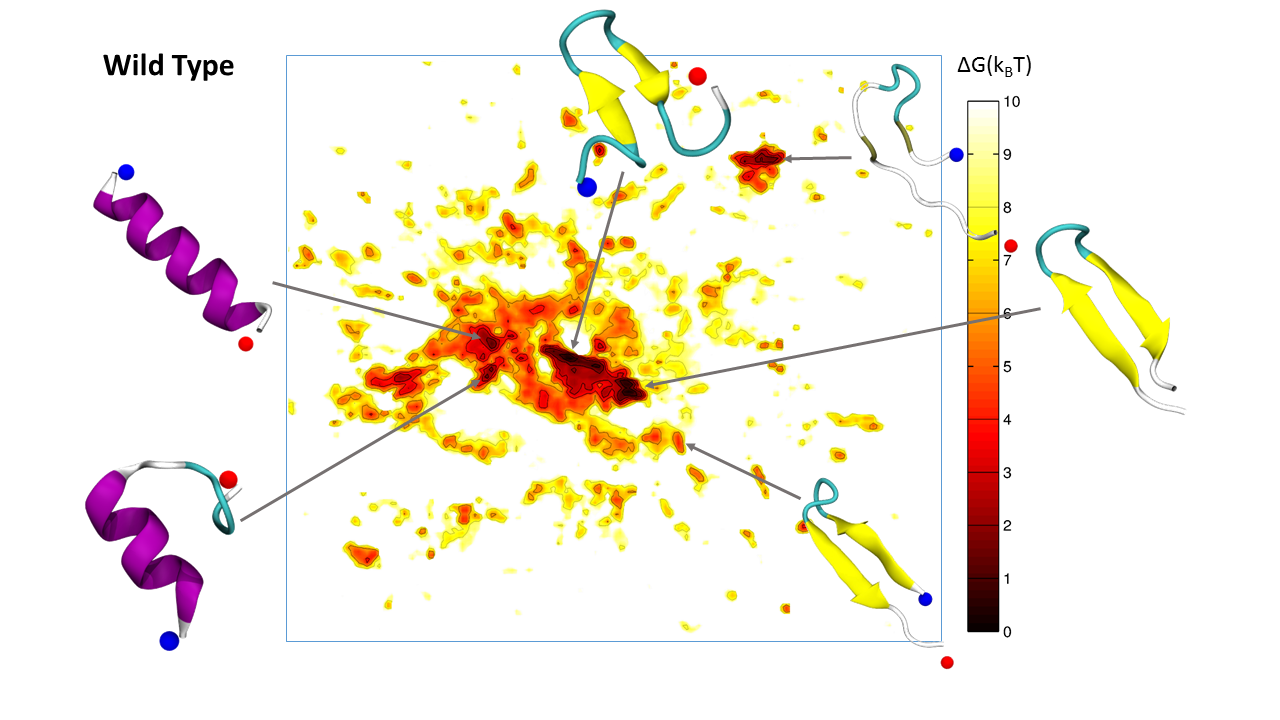}
 \caption{Free energy surface at 299~K for the wild type of the beta hairpin 
 protein as a function of the sketch-map coordinates.  The small inset figures 
 show  representative structures from each of the minima.}
 \label{fig:histo}
\end{figure}

\begin{figure}
 \centering
 \includegraphics[width=0.9\textwidth]{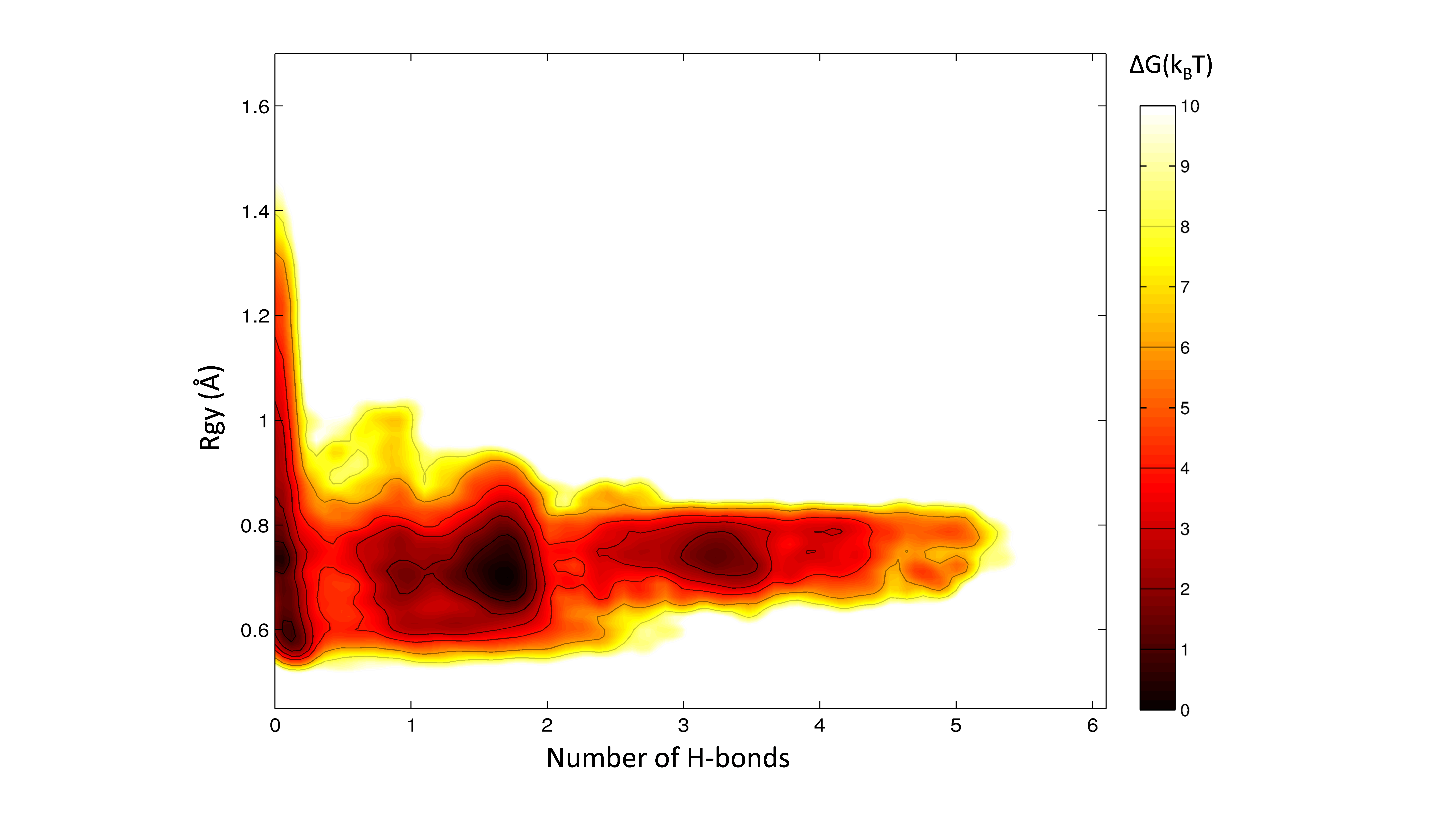}
 \caption{Free energy surface for the wild type protein studied in this work
 as a function of the number of hydrogen bonds that should be present
 in the folded configuration and the radius of gyration.  
The folded state is projected  on the right hand side of the figure, 
misfolded beta hairpin configurations appear in the centre of this figure while 
all other configurations, including alpha helical ones, are projected on the 
left hand side of this figure near to the y-axis. 
}
 \label{fig:cvs}
\end{figure}

Sketch-map coordinates were generated by selecting 1000 landmark points from our
wild-type trajectory using the staged algorithm  \cite{cluster-smap} with gamma equal 0.1 and wgamma
equal to 1. The set of Ramachandran angles for each protein configuration was
used for the high-dimensional, $\mathbf{X}$ vectors in equations \ref{eqn:smap}
and \ref{eqn:out-of-sample} rather than the position of all the atoms as 
by doing so we eliminate a large amount of redundant information
while still providing a good description of the 
variability in protein
structure. Optimal two-dimensional projections for each of the landmark protein
configurations were found by minimizing equation \ref{eqn:smap} with $\sigma=6$,
$A=8$, $B=8$, $a=2$ and $b=8$ \gt{using the optimization algorithm described in 
\cite{sketch-map}} .  Then, once this initial set of projections was
found, the remainder of the trajectory was projected into the sketch-map space
using equation \ref{eqn:out-of-sample}. \sout{so that} \gt{The free energy 
surface of the replica at 299.1 K shown in \ref{fig:histo} was then constructed 
by reweighting the histogram generated from our parallel tempering well 
tempered metadynamics simulations using the method described in 
\cite{max-reweighting}.}  There are a number of notable features in 
\ref{fig:histo}. 
Firstly, this figure shows that the free energy surface for the protein is very
rough.  There are many basins and each of these basins corresponds to a markedly
different protein configuration. This behaviour should be compared with that
observed in the free energy surface shown in \ref{fig:cvs}, in which the free
energy is shown as a function of the radius of gyration
and the degree to which the hydrogen bonds that would 
form in the beta hairpin have formed.  Importantly, the coordinate used to 
measure hydrogen bonding in this second figure is different to that used in the 
metadynamics simulations.  In the metadynamics our measure of the number of 
backbone atoms is agnostic and simply counts the number of backbone hydrogen 
bonds that have formed.  By contrast, for easy comparison with previous
literature, \cite{PTmetad, ale-urea, zhou-germain} in this figure we count only those 
hydrogen bonds that would be present in the final, folded beta-hairpin 
configuration.  

\ref{fig:cvs} shows a free energy surface that is considerably smoother than 
that shown in \ref{fig:histo}.  One now sees only three distinct
minima in the free energy landscape and many of the basins that were
visible in \ref{fig:histo} appear to have disappeared. It is therefore
apparent that when these commonly-used collective variables are used to analyse
the dynamics of the protein one can only really make judgements about the
relative stabilities of the folded (beta-hairpin) structure and the unfolded
state.  However, \ref{fig:histo} shows us that multiple configurations,
and perhaps more pertinently multiple energetic basins, make up this
``unfolded''
state.  There is incomplete agreement in the literature about what structures
together comprise the unfolded ensemble for this protein.  While some simulation
studies suggest that there is no significant alpha-helical content \cite{zhou-germain} other see a
substantial number of helical configurations\cite{best-mittal,garc-sanb,best}.  The free energy surface in 
\ref{fig:histo} seems to suggest that alpha helical configurations
are present and that they have free energies that are comparable with those of 
the beta-hairpin.

Another contentious issue in the literature on this protein concerns the
various misfolded configurations that the protein can adopt.  Bonomi \emph{et
al.} \cite{bonomi} suggested the existence of one misfolded state.  Meanwhile, Best \emph{et
al.} \cite{best} found multiple misfolded configurations, which included Bonomi's. In our
free energy surface we see many of the misfolded configurations that were
observed in these studies. We do not see all because in these other studies
different force fields were used, which perhaps stabilize different
configurations.  In our work we used the AMBER99SB-ILDN* \cite{amber-ildn} because it has been shown
to reproduce experimental observations for small peptides and proteins well \cite{lindorff}.  

\begin{figure}
 \centering
 \includegraphics[width=0.9\textwidth]{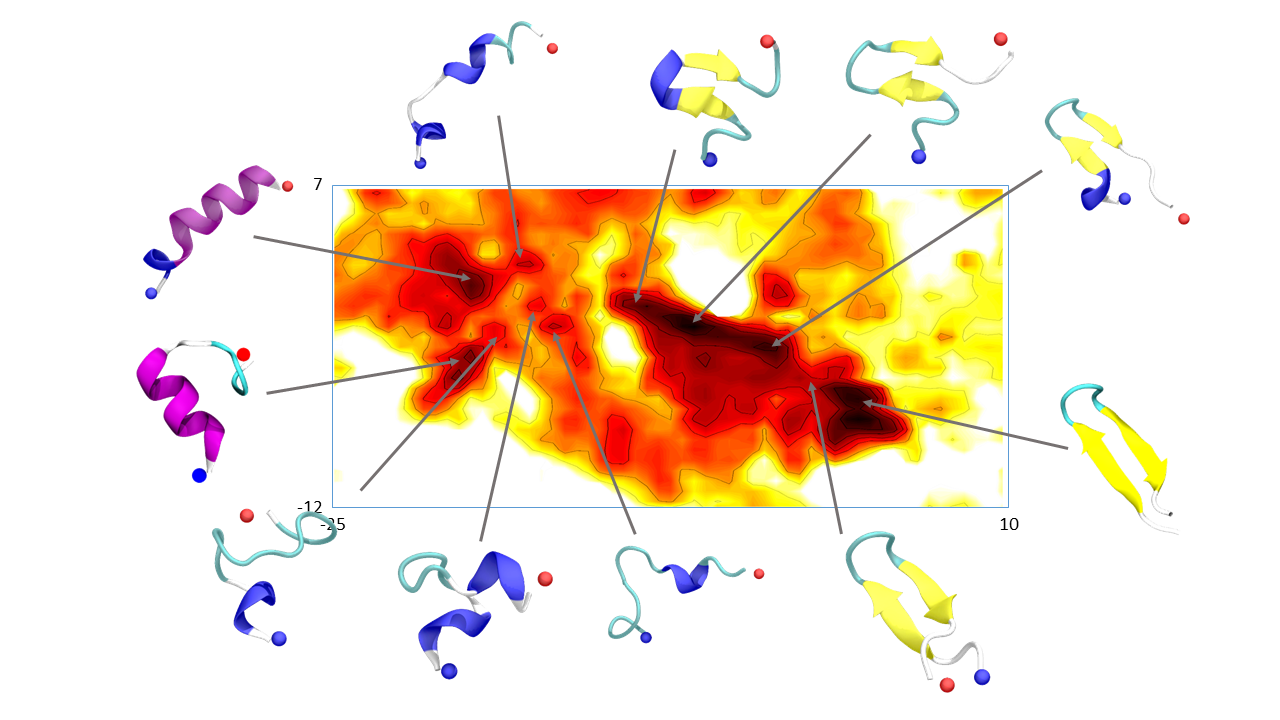}
 \caption{Figure showing the region of the sketch-map projection of the free 
energy surface
 containing the beta hairpin and alpha helical configurations.  
Representative configurations for each 
 of the minima in this region are shown.}
 \label{fig:pathway}
\end{figure}

\ref{fig:pathway} shows the region of the sketch-map free energy surface around 
the
alpha helical and beta hairpin minima in more detail.  This figure shows clearly that 
there are many minima in the free energy landscape in this region corresponding to 
structures with various degrees of alpha helical and beta-sheet-like character.  
The most 
stable configuration is the folded beta hairpin.  However, there are a number of partially
folded hairpins that lie relatively close in energy.  These configurations have a free energy
that is only slightly lower than that of a perfect alpha helix. Most 
importantly, however, a comparison between figures 1 and 2 clearly shows that 
variables such as the radius of gyration or the number of 
native hydrogen bonds, which measure the degree to which the structure 
resembles that in the folded state, give an incomplete picture of the unfolded 
state.  Alpha-helical configurations appear in a separate basin from the 
misfolded configurations in \ref{fig:histo} but are combined in a single basin 
in \ref{fig:cvs}.   

It is tempting to look for the transition pathways between configurations 
in these sketch-map coordinates.  Care should be taken when doing so, however, 
as firstly no dynamical information  was used in the construction of these 
coordinates.  At best structures appear close together in the projection because 
they are relatively close together in the high-dimensional space.  This does 
not necessarily mean that the system will rapidly inter-convert between these 
configurations - two structures that appear close together may well be 
separated by a substantial energy barrier.  Secondly, we have shown 
\cite{fieldcvs} that continuous paths in the high-dimensional space can appear 
to be discontinuous when projected using sketch-map.

\begin{figure}
 \centering
 \includegraphics[width=0.9\textwidth]{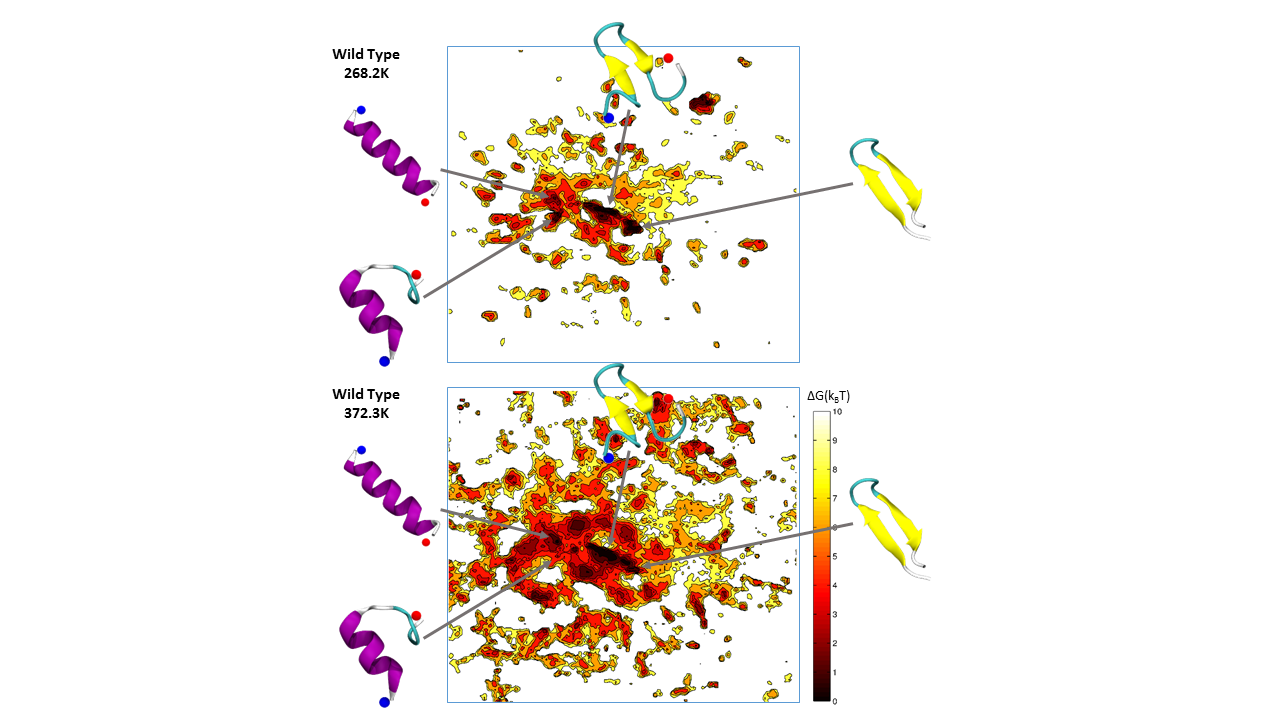}
 \caption{Free energy surfaces for the wild type protein examined in this work 
 as a function of the sketch-map coordinates at temperatures of 268.2~K and 
372.3~K.
 These surfaces were generated by projecting the trajectories for the replicas 
at these 
 temperatures using a set of landmarks extracted from the trajectory at 299.2~K. 
 As such 
 the various basins in these landscapes can be compared directly with those in  
 \ref{fig:histo}. The figure shows that at low temperatures the system explores 
a small 
 part of configuration space and is largely confined to the folded states.  At 
higher temperatures
 the system explores more of configuration space including a wide variety of 
 unfolded structures.  However, this free energy surface tells us that the 
system will spend a considerable portion of its time in folded configuration 
even at these relatively high temperatures.}
 \label{fig:temps}
\end{figure}

 Plotting the free energy surface as a 
function of the sketch-map variables at a variety of different 
temperatures using a single reference map obtained at an intermediate 
temperature is an interesting exercise. When this is done for the 
trajectories in this work we observe results that are similar to those obtained 
in a recent work on Lennard-Jones clusters \cite{cluster-smap}.
 \ref{fig:temps} shows that at low temperatures the system is
confined to the low-energy folded states - the alpha helical and beta sheet
configurations.  By contrast at higher temperatures the system is free to
explore a much wider portion of configuration space including many high energy
configurations.  
It is interesting to note that basins corresponding to the hairpin, the mis-folded hairpin and the 
helix, are present at all temperatures. In other words, the system spends a 
substantial amount of time in folded states even when it is at a temperature 
of 373 K.
The apparent stabilities of these folded configurations
can be rationalized by noting that 
the basins corresponding to each of them are  wider in 
the higher temperature free energy surface. It would 
seem then that increases in entropy that 
occur because the system fluctuates more wildly about the equilibrium 
structures at higher temperatures serve to  stabilize the 
folded structure even when the temperature  is high. 
This should be contrasted with the case of small clusters \cite{cluster-smap}, 
where the enthalpic  minima appear to be  very 
rigid.  In these systems structural fluctuations about these minima 
appear not to increase strongly with temperature.  Consequently, the 
weights of these enthalpic basins becomes  completely 
irrelevant  when the temperature is raised.

\subsection{Trpzip4 and D46A mutations}

\begin{figure}
 \centering
 \includegraphics[width=0.9\textwidth]{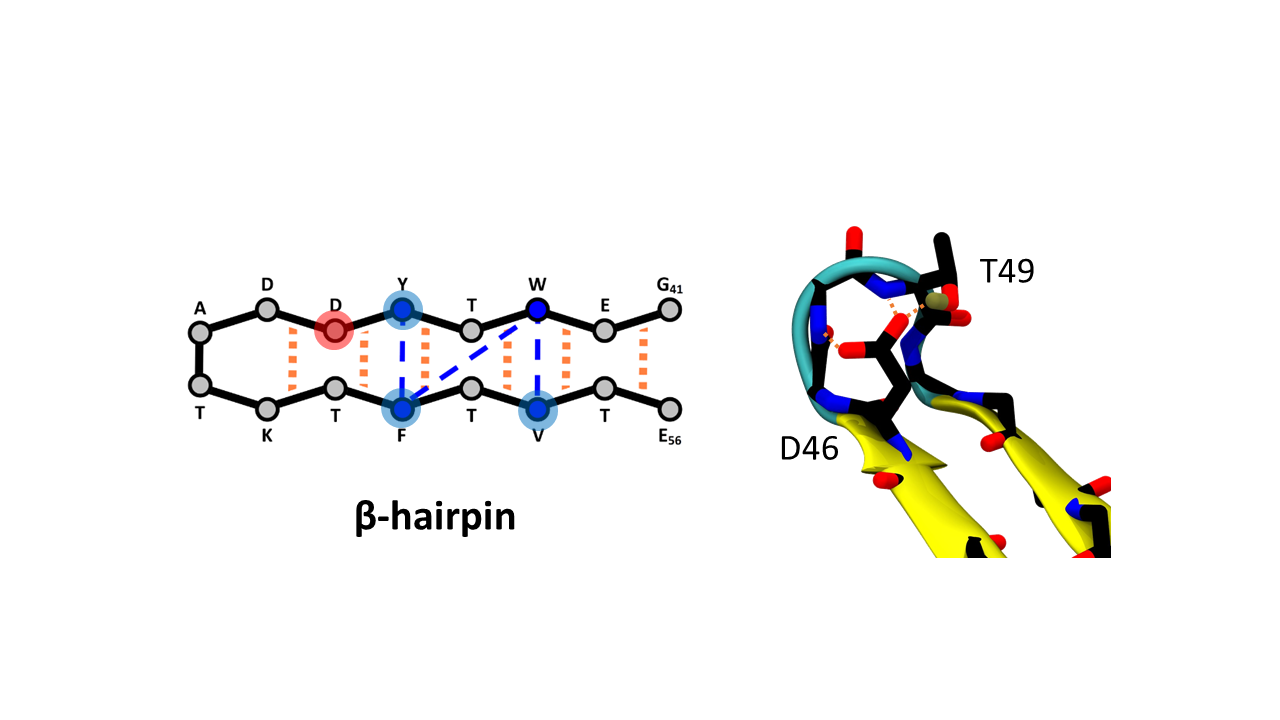}
 \caption{Figure illustrating the mutations we have studied in this work.
 The left panel shows a schematic of the folded configuration.  The red dashed
 lines indicate the various hydrogen bonds that form between backbone groups in 
the folded state while the blue dashed lines illustrate the favourable hydrophobic 
interactions that serve to stabilize the folded state.  In the first mutation we 
examined the residues highlighted in blue were changed to tryptophans. In the second 
mutation the aspartate residue highlighted in red was changed to an alanine. As shown 
in the right panel of the figure this aspartate (D46) forms a hydrogen bond to residue 
T49 in the wild type, which serves to stabilise the loop region for the folded beta 
hairpin. Alanine cannot form this hydrogen bond and thus this mutation destabilizes 
the folded state.}
 \label{fig:mutations}
\end{figure}

\begin{figure}
 \centering
 \includegraphics[width=0.9\textwidth]{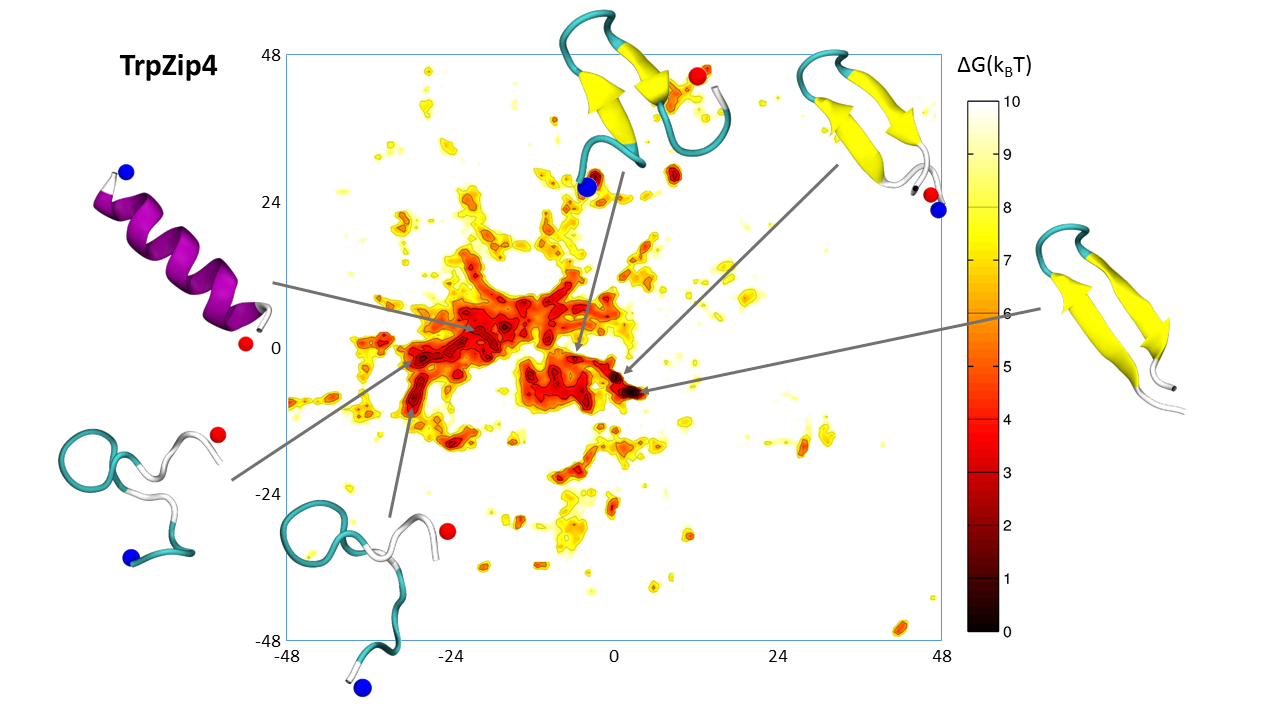}
 \caption{Free energy surface at 299~K for a mutant protein in which residues 
45, 52 and
 54 have been changed to tryptophans. By comparing this landscape with that of 
the 
 wild type in \ref{fig:histo} it becomes apparent that this mutation 
serves to stabilize
 the folded beta hairpin configuration.}
 \label{fig:mut1}
\end{figure}

\begin{figure}
 \centering
 \includegraphics[width=0.9\textwidth]{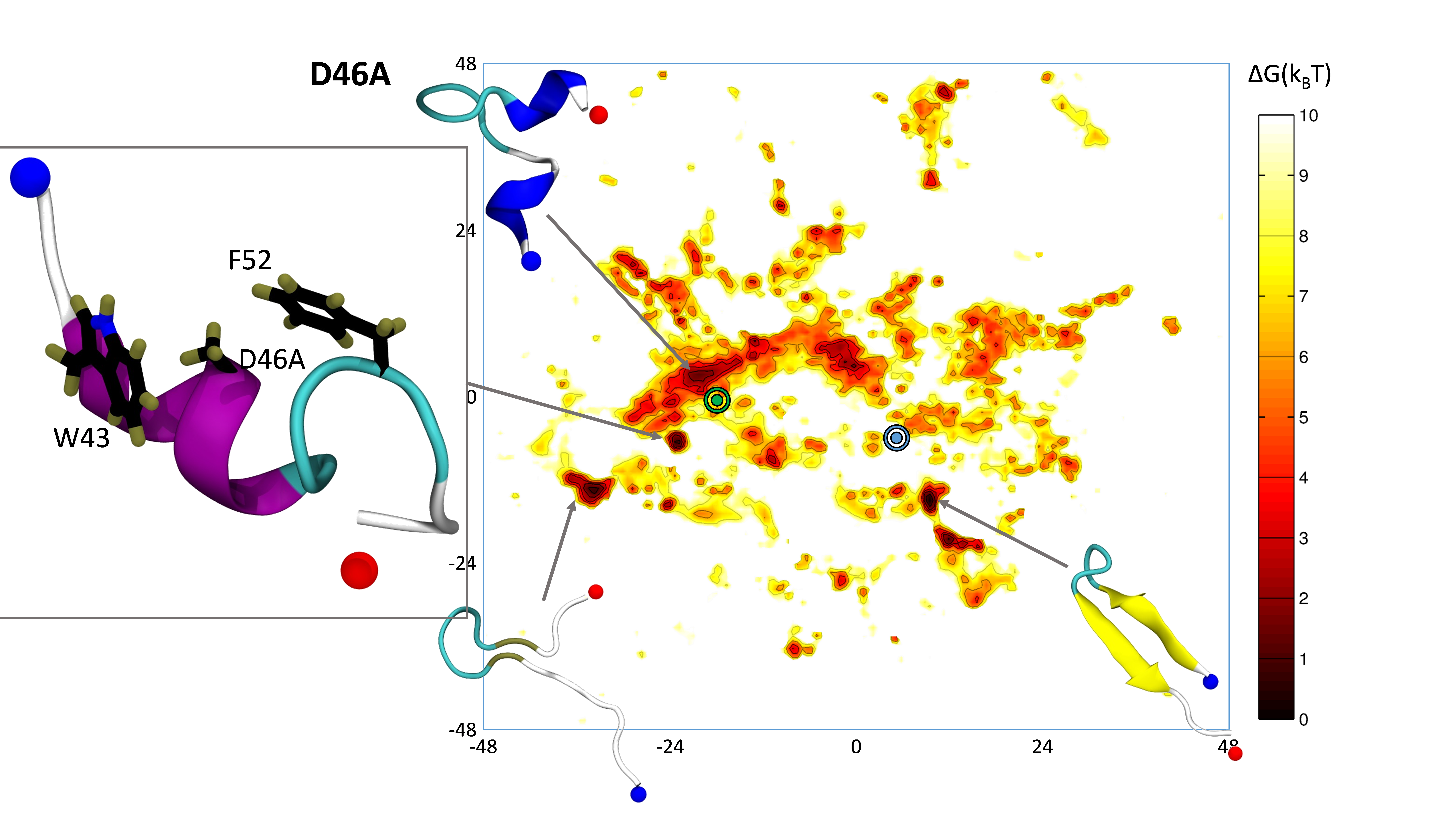}
 \caption{Free energy surface at 299~K for a mutant protein in which residue 46
 has been changed into an alanine. \avol{The blue and the green markpoints show the 
 position of the folded and the linear alpha helix configurations respectively as observed 
 in the wild type FES.} By comparing this landscape with that of the 
 wild type in \ref{fig:histo} it becomes apparent that this mutation 
 serves to destabilize the folded beta hairpin configuration as well as the 
linear alpha helix. As discussed in the text we see from our simulation that 
the alanine on residue 46, as well as not forming the hydrogen bonds that 
stabilise the beta hairpin, can also form the hydrophobic contacts shown in 
the left panel of this figure. This serves to destabilize the linear 
alpha helix.}
 \label{fig:mut2}
\end{figure}

It has been suggested that protein energy landscapes always have minima
corresponding to the various secondary structural elements as these features in
the energy landscape emerge as a result of interactions between the backbone
atoms that are unaffected by the amino acid sequence \cite{backbone,spectrum}. In this view the amino
acid sequence only perturbs the energies of this universal library of 
secondary structure motifs.  
As a result the sequence, rather than controlling the shape
of the funnel that leads to the folded state \cite{funnel}, serves only to select a folded
state from a sm{\"o}rg\aa{}sbord of allowable protein configurations.  
The sketch-map
coordinates discussed in the previous section provide us with an interesting
opportunity to examine this kind of hypothesis.  By projecting the free energy
surface for a mutant protein using the sketch-map coordinates that we
generated in the previous section for the wild-type protein we can do a
basin-by-basin comparison of the free energy landscapes for two proteins with
different amino acid sequences.  In other words we can examine how the free
energy for each of the individual basins in the energy landscape is perturbed by
the mutation and can thus perhaps provide more detailed insight than 
observing simply that one mutation
stabilizes the folded state while the other destabilizes it.

With the above experiment in mind we ran
parallel-tempering-well-tempered-metadynamics simulations similar to those
described at the start of the previous section on two mutations of the
beta-hairpin protein.  In the first of these mutants we changed the tyrosine,
phenylalanine and Valine groups on residues 45, 52 and 54 to tryptophans.  The residues 
we mutated are shown in blue in the left panel of \ref{fig:mutations}.  This
figure shows that the mutated residues participate in a number of hydrophobic interactions
that stabilize the core of the structure.  Hence, by making these already 
hydrophobic residues even more hydrophobic we would expect to further stabilize 
the folded beta-hairpin configuration \cite{noy-keasar,jura-bolh}, which is 
precisely what is observed in \ref{fig:mut1}.  The minima corresponding 
to the beta-hairpin configuration becomes deeper as do the satellite basins 
around it that correspond to the various partially unfolded
configurations of this structure. 

\ref{fig:mut1} shows that the free energy difference between the 
perfect-alpha helix structure and the folded beta hairpin is larger for this 
mutated protein than it is in the wild type.  Even so the alpha-helix still has 
a reasonably low free energy so the system would still expect to spend a 
considerable fraction of its time in alpha helical configurations. The reason 
for this is probably connected to the fact that alpha helices are held 
together by backbone hydrogen bonds between carboxylate and amine groups that 
are four residues apart on the protein chain. A mutation of the protein would 
not be expected to change the strength of these particular interactions 
significantly.  A significant change in the hydrophobicity of the amino acid 
side chains might, by contrast, act to destabilize the alpha helix because, in 
an alpha helix, all the side chains are exposed to the solvent.  It would seem 
from \ref{fig:mut1}, however, that making hydrophobic residues slightly 
more hydrophobic does not perturb the free energy landscape greatly 
enough to bring about this change.

In the second mutation we substituted the aspartate on position 46, which is 
shown in red in the left panel of \ref{fig:mutations}, for an
alanine.  As shown in the right panel of \ref{fig:mutations} in the 
wild-type this aspartate stabilises the loop region of the folded state by 
forming hydrogen bonds both to the backbone amide nitrogens of those residues in 
the loop and to the side-chain of threonine 49.  Alanine will 
not
form these hydrogen bonds,  which explains why 
there is no longer a basin corresponding to the beta-hairpin structure or any 
satellite basins corresponding to partially-unfolded beta hairpins in the free 
energy landscape for this mutant (see \ref{fig:mut2}). 
Interestingly, this mutation also appears to affect the stability of the 
perfect alpha-helix.  The linear alpha-helical configuration that appears in 
\ref{fig:histo} and \ref{fig:mut1} is not shown in  
\ref{fig:mut2} because the system did not visit this particular configuration 
in our parallel tempering simulations on the D46A mutant.  This fact is 
evidenced in the free energy surfaces shown in figures 1, 6 and 7. 
In the first two of these free energy 
surfaces the linear alpha helix is shown to correspond to a rather distinctive 
sausage-shaped basin in the free energy landscape, which is no longer visible in 
\ref{fig:mut2}.  To be clear, it is not that this mutant does not visit 
alpha helices at all it is rather that the alpha helical configurations it does 
visit are folded back on themselves (see the left panel of  
\ref{fig:mut2}). One explanation for this is that by substituting the aspartate 
on residue 46 with an alanine we have made a formerly hydrophilic residue 
hydrophobic. As discussed previously, this would destabilize the alpha helix as 
in this configuration all the side chains are exposed to the solvent. In 
addition, the left panel of \ref{fig:mut2} show that this mutated alanine 
group forms two hydrophobic contacts that stabilise a bent alpha helix over 
the linear one.

As discussed previously the global minimum for both the 
wild-type and the TrpZip4 mutant is a beta hairpin configuration. There is 
no basin corresponding to this structure in the D46A free energy surface for 
the reasons discussed in the previous paragraph.  However, this is not to say 
that beta hairpins do not form for this particular amino acid sequence.  There 
is a prominent basin that corresponds to a structure in which an anti parallel 
beta hairpin has formed and in which the loop is between residues 44 and 47 in 
the bottom right hand corner of the free energy surface shown in 
\ref{fig:mut2}. This structure is favourable because within it there are 
numerous hydrophobic contacts.  Furthermore, it appears in the free energy 
landscapes for both the wild-type and TrpZip4 mutant.  However, in those 
proteins 
the feature in the free energy landscape corresponding to this structure is 
considerably less prominent because this structure contains fewer 
hydrophobic contacts than the perfect beta hairpin.  The reason it is so 
prominent for the D46A is almost certainly connected to the fact that it is 
unfavourable to form the loop between residues 47 and 51 because the (mutated) 
alanine group on residue 46 cannot form the prerequisite hydrogen bonds.

\section{Conclusion}

The sketch-map coordinates that have been used to 
analyse the long MD trajectories generated in this study
have shown themselves to be a useful tool.  
When the free energy surface is displayed as a function of these bespoke
coordinates one can see the wide range of configurations adopted by a 
particular chemical system. For proteins this means that one can use
sketch-map coordinates to view, 
basin by basin, how a change in the conditions or a change in the amino acid 
sequence affects the free energy landscape of a protein.  This is useful as 
evidence keeps emerging that denaturants such as urea \cite{ale-urea} work by 
changing the relative free energy of unfolded configurations. Similarly 
a better understanding of how the free energy landscapes of proteins change in 
response to experimental conditions will allow us to better understand phenomena such as the 
Hoffmeister series \cite{hoffmeister} or even how changing the forcefield 
parameters changes the way phase space is sampled.  Lastly, there is increasing 
evidence 
that so called intrinsically disordered proteins play an important role in 
biology. Many of these proteins undergo folding and binding at the same time 
and hence untangling their modus operandi requires a detailed view on how 
free energies of folding are perturbed by binding and vice versa.

Our results suggest that mutations can stabilize or destabilize 
particular configurations of the peptide by either allowing it to form 
new energetically-favourable contacts or by preventing them from forming 
particular native contacts. It seems reasonable to suppose that this 
may well be valid for other small, isolated proteins so our results would 
appear to support backbone theories \cite{backbone,spectrum} of protein folding 
more than they support those theories based on the amino acids sequence 
sculpting the folding funnel \cite{wolynes}. However, in a bigger protein 
regions of secondary structure are likely to be stabilized by additional 
interactions that will all serve to ensure that the 
protein adopts its biologically-active, tertiary structure.  In these 
proteins the tertiary structure is thus less likely to be affected by small 
(point)} mutations.  As such further study is thus certainly 
required if we are to determine the effect mutations have on the folding 
landscape in general terms.

We were able to come to these conclusions because sketch-map coordinates allow 
one to
perform an unbiased analysis of the simulation results.  Additionally, equation
\ref{eqn:out-of-sample} ensures that we can calculate projections easily for any
vector of 30 torsional angles.  We can even use these coordinates to enhance 
sampling and to thus drive the system to explore high energy parts of 
configuration space \cite{fieldcvs}. All the evidence we have accumulated thus 
far
further suggests that this procedure will produce sensible projections.  We thus
have coordinates that can be used for a number of different amino acid 
sequences
and which are not constructed based on
previous knowledge on the structure of the folded state.  In short, we are in a
position where we can compare data collected from multiple trajectories and
where our previous understanding does not limit our analysis.  As such there is
thus a greater chance of discovering unexpected or surprising results.

\section{Acknowledgements}

The authors would like to thank CSCS for computer time and also acknowledge 
funding from the European Union (Grant ERC-2009-AdG-24707). AA was supported by
an EMBO long-term fellowship.

\providecommand{\latin}[1]{#1}
\makeatletter
\providecommand{\doi}
  {\begingroup\let\do\@makeother\dospecials
  \catcode`\{=1 \catcode`\}=2 \doi@aux}
\providecommand{\doi@aux}[1]{\endgroup\texttt{#1}}
\makeatother
\providecommand*\mcitethebibliography{\thebibliography}
\csname @ifundefined\endcsname{endmcitethebibliography}
  {\let\endmcitethebibliography\endthebibliography}{}


\begin{mcitethebibliography}{46}
\providecommand*\natexlab[1]{#1}
\providecommand*\mciteSetBstSublistMode[1]{}
\providecommand*\mciteSetBstMaxWidthForm[2]{}
\providecommand*\mciteBstWouldAddEndPuncttrue
  {\def\EndOfBibitem{\unskip.}}
\providecommand*\mciteBstWouldAddEndPunctfalse
  {\let\EndOfBibitem\relax}
\providecommand*\mciteSetBstMidEndSepPunct[3]{}
\providecommand*\mciteSetBstSublistLabelBeginEnd[3]{}
\providecommand*\EndOfBibitem{}
\mciteSetBstSublistMode{f}
\mciteSetBstMaxWidthForm{subitem}{(\alph{mcitesubitemcount})}
\mciteSetBstSublistLabelBeginEnd
  {\mcitemaxwidthsubitemform\space}
  {\relax}
  {\relax}

\bibitem[Lodish \latin{et~al.}(2003)Lodish, Berk, Matsudaira, Kaiser, Krieger,
  Scott, Zipursky, and Darnell]{bio-book}
Lodish,~H.; Berk,~A.; Matsudaira,~P.; Kaiser,~C.~A.; Krieger,~M.; Scott,~M.~P.;
  Zipursky,~S.~L.; Darnell,~J. \emph{{Molecular Cell Biology}}, {Fifth} ed.; {W
  H Freeman}: {New York}, 2003\relax
\mciteBstWouldAddEndPuncttrue
\mciteSetBstMidEndSepPunct{\mcitedefaultmidpunct}
{\mcitedefaultendpunct}{\mcitedefaultseppunct}\relax
\EndOfBibitem
\bibitem[Dunker \latin{et~al.}(2008)Dunker, Silman, Uversky, and
  Sussman]{signal}
Dunker,~A.~K.; Silman,~I.; Uversky,~V.~N.; Sussman,~J.~L. \emph{Curr. Opin.
  Struct. Biol.} \textbf{2008}, \emph{18}, 756--764\relax
\mciteBstWouldAddEndPuncttrue
\mciteSetBstMidEndSepPunct{\mcitedefaultmidpunct}
{\mcitedefaultendpunct}{\mcitedefaultseppunct}\relax
\EndOfBibitem
\bibitem[Christopoulos(2002)]{allosteric}
Christopoulos,~A. \emph{Nat. Rev. Drug Discovery} \textbf{2002}, \emph{1},
  198--210\relax
\mciteBstWouldAddEndPuncttrue
\mciteSetBstMidEndSepPunct{\mcitedefaultmidpunct}
{\mcitedefaultendpunct}{\mcitedefaultseppunct}\relax
\EndOfBibitem
\bibitem[Dyson and Wright(2005)Dyson, and Wright]{idp}
Dyson,~H.~J.; Wright,~P.~E. \emph{Nat. Rev. Mol. Cell Biol.} \textbf{2005},
  \emph{6}, 197--208\relax
\mciteBstWouldAddEndPuncttrue
\mciteSetBstMidEndSepPunct{\mcitedefaultmidpunct}
{\mcitedefaultendpunct}{\mcitedefaultseppunct}\relax
\EndOfBibitem
\bibitem[Shaw \latin{et~al.}(2010)Shaw, Maragakis, Lindorff-Larsen, Piana,
  Dror, Eastwood, Bank, Jumper, Salmon, Shan, and Wriggers]{long-md-simulation}
Shaw,~D.~E.; Maragakis,~P.; Lindorff-Larsen,~K.; Piana,~S.; Dror,~R.~O.;
  Eastwood,~M.~P.; Bank,~J.~A.; Jumper,~J.~M.; Salmon,~J.~K.; Shan,~Y.;
  Wriggers,~W. Atomic-Level Characterization of the Structural Dynamics of
  Proteins. \emph{Science} \textbf{2010}, \emph{330}, 341--346\relax
\mciteBstWouldAddEndPuncttrue
\mciteSetBstMidEndSepPunct{\mcitedefaultmidpunct}
{\mcitedefaultendpunct}{\mcitedefaultseppunct}\relax
\EndOfBibitem
\bibitem[Li and Cirino(2014)Li, and Cirino]{mutagenesis}
Li,~Y.; Cirino,~P.~C. Recent advances in engineering proteins for biocatalysis.
  \emph{Biotechnol. Bioeng.} \textbf{2014}, \emph{111}, 1273--1287\relax
\mciteBstWouldAddEndPuncttrue
\mciteSetBstMidEndSepPunct{\mcitedefaultmidpunct}
{\mcitedefaultendpunct}{\mcitedefaultseppunct}\relax
\EndOfBibitem
\bibitem[Bouvignies \latin{et~al.}({2011})Bouvignies, Vallurupalli, Hansen,
  Correia, Lange, Bah, Vernon, Dahlquist, Baker, and Kay]{trans-lyzo}
Bouvignies,~G.; Vallurupalli,~P.; Hansen,~D.~F.; Correia,~B.~E.; Lange,~O.;
  Bah,~A.; Vernon,~R.~M.; Dahlquist,~F.~W.; Baker,~D.; Kay,~L.~E. {Solution
  structure of a minor and transiently formed state of a T4 lysozyme mutant}.
  \emph{{Nature}} \textbf{{2011}}, \emph{{477}}, {111--U134}\relax
\mciteBstWouldAddEndPuncttrue
\mciteSetBstMidEndSepPunct{\mcitedefaultmidpunct}
{\mcitedefaultendpunct}{\mcitedefaultseppunct}\relax
\EndOfBibitem
\bibitem[Bussi \latin{et~al.}(2006)Bussi, Gervasio, Laio, and
  Parrinello]{PTmetad}
Bussi,~G.; Gervasio,~F.~L.; Laio,~A.; Parrinello,~M. Free-Energy Landscape for
  $\beta$ Hairpin Folding from Combined Parallel Tempering and Metadynamics.
  \emph{J. Chem. Am. Soc.} \textbf{2006}, \emph{128}, 13435--13441, PMID:
  17031956\relax
\mciteBstWouldAddEndPuncttrue
\mciteSetBstMidEndSepPunct{\mcitedefaultmidpunct}
{\mcitedefaultendpunct}{\mcitedefaultseppunct}\relax
\EndOfBibitem
\bibitem[Gnanakaran \latin{et~al.}(2003)Gnanakaran, Nymeyer, Portman,
  Sanbonmatsu, and Garcia]{gnanakaran}
Gnanakaran,~S.; Nymeyer,~H.; Portman,~J.; Sanbonmatsu,~K.~Y.; Garcia,~A.~E.
  Peptide folding simulations. \emph{Curr. Opin. Struct. Biol.} \textbf{2003},
  \emph{13}, 168 -- 174\relax
\mciteBstWouldAddEndPuncttrue
\mciteSetBstMidEndSepPunct{\mcitedefaultmidpunct}
{\mcitedefaultendpunct}{\mcitedefaultseppunct}\relax
\EndOfBibitem
\bibitem[Weinstock \latin{et~al.}(2007)Weinstock, Narayanan, Felts, Andrec,
  Levy, Wu, and Baum]{weinstock}
Weinstock,~D.~S.; Narayanan,~C.; Felts,~A.~K.; Andrec,~M.; Levy,~R.~M.;
  Wu,~K.-P.; Baum,~J. Distinguishing among Structural Ensembles of the GB1
  Peptide: REMD Simulations and NMR Experiments. \emph{J. Am. Chem. Soc.}
  \textbf{2007}, \emph{129}, 4858--4859, PMID: 17402734\relax
\mciteBstWouldAddEndPuncttrue
\mciteSetBstMidEndSepPunct{\mcitedefaultmidpunct}
{\mcitedefaultendpunct}{\mcitedefaultseppunct}\relax
\EndOfBibitem
\bibitem[Deighan \latin{et~al.}(2012)Deighan, Bonomi, and Pfaendtner]{deighan}
Deighan,~M.; Bonomi,~M.; Pfaendtner,~J. Efficient Simulation of Explicitly
  Solvated Proteins in the Well-Tempered Ensemble. \emph{J. Chem. Theory
  Comput.} \textbf{2012}, \emph{8}, 2189--2192\relax
\mciteBstWouldAddEndPuncttrue
\mciteSetBstMidEndSepPunct{\mcitedefaultmidpunct}
{\mcitedefaultendpunct}{\mcitedefaultseppunct}\relax
\EndOfBibitem
\bibitem[Zuckerman(2011)]{zuckerman}
Zuckerman,~D.~M. Equilibrium Sampling in Biomolecular Simulations. \emph{Annual
  Review of Biophysics} \textbf{2011}, \emph{40}, 41--62, PMID: 21370970\relax
\mciteBstWouldAddEndPuncttrue
\mciteSetBstMidEndSepPunct{\mcitedefaultmidpunct}
{\mcitedefaultendpunct}{\mcitedefaultseppunct}\relax
\EndOfBibitem
\bibitem[Ceriotti \latin{et~al.}(2011)Ceriotti, Tribello, and
  Parrinello]{sketch-map}
Ceriotti,~M.; Tribello,~G.~A.; Parrinello,~M. Simplifying the representation of
  complex free-energy landscapes using sketch-map. \emph{Proc. Natl. Acad. Sci.
  USA} \textbf{2011}, \emph{108}, 13023--13029\relax
\mciteBstWouldAddEndPuncttrue
\mciteSetBstMidEndSepPunct{\mcitedefaultmidpunct}
{\mcitedefaultendpunct}{\mcitedefaultseppunct}\relax
\EndOfBibitem
\bibitem[Tenenbaum \latin{et~al.}(2000)Tenenbaum, Silva, and Langford]{isomap}
Tenenbaum,~J.~B.; Silva,~V.~d.; Langford,~J.~C. A Global Geometric Framework
  for Nonlinear Dimensionality Reduction. \emph{Science} \textbf{2000},
  \emph{290}, 2319--2323\relax
\mciteBstWouldAddEndPuncttrue
\mciteSetBstMidEndSepPunct{\mcitedefaultmidpunct}
{\mcitedefaultendpunct}{\mcitedefaultseppunct}\relax
\EndOfBibitem
\bibitem[Roweis and Saul(2000)Roweis, and Saul]{lle}
Roweis,~S.~T.; Saul,~L.~K. Nonlinear Dimensionality Reduction by Locally Linear
  Embedding. \emph{Science} \textbf{2000}, \emph{290}, 2323--2326\relax
\mciteBstWouldAddEndPuncttrue
\mciteSetBstMidEndSepPunct{\mcitedefaultmidpunct}
{\mcitedefaultendpunct}{\mcitedefaultseppunct}\relax
\EndOfBibitem
\bibitem[Coifman \latin{et~al.}(2005)Coifman, Lafon, Lee, Maggioni, Nadler,
  Warner, and Zucker]{diffmap-1}
Coifman,~R.~R.; Lafon,~S.; Lee,~A.~B.; Maggioni,~M.; Nadler,~B.; Warner,~F.;
  Zucker,~S.~W. Geometric diffusions as a tool for harmonic analysis and
  structure definition of data: Multiscale methods. \emph{Proc. Natl. Acad.
  Sci. USA of the United States of America} \textbf{2005}, \emph{102},
  7432--7437\relax
\mciteBstWouldAddEndPuncttrue
\mciteSetBstMidEndSepPunct{\mcitedefaultmidpunct}
{\mcitedefaultendpunct}{\mcitedefaultseppunct}\relax
\EndOfBibitem
\bibitem[Coifman and Lafon(2006)Coifman, and Lafon]{diffmap-2}
Coifman,~R.~R.; Lafon,~S. Diffusion maps. \emph{Appl. Comput. Harmon. Anal.}
  \textbf{2006}, \emph{21}, 5 -- 30\relax
\mciteBstWouldAddEndPuncttrue
\mciteSetBstMidEndSepPunct{\mcitedefaultmidpunct}
{\mcitedefaultendpunct}{\mcitedefaultseppunct}\relax
\EndOfBibitem
\bibitem[Belkin and Niyogi(2003)Belkin, and Niyogi]{diffmap-3}
Belkin,~M.; Niyogi,~P. Laplacian Eigenmaps for Dimensionality Reduction and
  Data Representation. \emph{Neural Comput.} \textbf{2003}, \emph{15},
  1373--1396\relax
\mciteBstWouldAddEndPuncttrue
\mciteSetBstMidEndSepPunct{\mcitedefaultmidpunct}
{\mcitedefaultendpunct}{\mcitedefaultseppunct}\relax
\EndOfBibitem
\bibitem[Tribello \latin{et~al.}(2012)Tribello, Ceriotti, and
  Parrinello]{fieldcvs}
Tribello,~G.~A.; Ceriotti,~M.; Parrinello,~M. Using sketch-map coordinates to
  analyze and bias molecular dynamics simulations. \emph{Proc. Natl. Acad. Sci.
  USA} \textbf{2012}, \emph{109}, 5196--5201\relax
\mciteBstWouldAddEndPuncttrue
\mciteSetBstMidEndSepPunct{\mcitedefaultmidpunct}
{\mcitedefaultendpunct}{\mcitedefaultseppunct}\relax
\EndOfBibitem
\bibitem[Cox and Cox(1994)Cox, and Cox]{mds}
Cox,~T.~F.; Cox,~M. A.~A. \emph{Multidimensional Scaling}; London: Chapman and
  Hall, 1994\relax
\mciteBstWouldAddEndPuncttrue
\mciteSetBstMidEndSepPunct{\mcitedefaultmidpunct}
{\mcitedefaultendpunct}{\mcitedefaultseppunct}\relax
\EndOfBibitem
\bibitem[Ceriotti \latin{et~al.}(2013)Ceriotti, Tribello, and
  Parrinello]{cluster-smap}
Ceriotti,~M.; Tribello,~G.~A.; Parrinello,~M. Demonstrating the Transferability
  and the Descriptive Power of Sketch-Map. \emph{J. Chem. Theory Comput.}
  \textbf{2013}, \emph{9}, 1521--1532\relax
\mciteBstWouldAddEndPuncttrue
\mciteSetBstMidEndSepPunct{\mcitedefaultmidpunct}
{\mcitedefaultendpunct}{\mcitedefaultseppunct}\relax
\EndOfBibitem
\bibitem[Blanco \latin{et~al.}(1994)Blanco, Rivas, and Serrano]{hairpin1}
Blanco,~F.~J.; Rivas,~G.; Serrano,~L. \emph{Nat. Struct. Biol.} \textbf{1994},
  \emph{1}, 584--590\relax
\mciteBstWouldAddEndPuncttrue
\mciteSetBstMidEndSepPunct{\mcitedefaultmidpunct}
{\mcitedefaultendpunct}{\mcitedefaultseppunct}\relax
\EndOfBibitem
\bibitem[Hughes and Waters(2006)Hughes, and Waters]{hairpin2}
Hughes,~R.; Waters,~M. \emph{Curr. Opin. Struct. Biol.} \textbf{2006},
  \emph{16}, 514--524\relax
\mciteBstWouldAddEndPuncttrue
\mciteSetBstMidEndSepPunct{\mcitedefaultmidpunct}
{\mcitedefaultendpunct}{\mcitedefaultseppunct}\relax
\EndOfBibitem
\bibitem[Mu{\~n}oz \latin{et~al.}(1997)Mu{\~n}oz, Thompson, Hofrichter, and
  Easton]{hairpin3}
Mu{\~n}oz,~V.; Thompson,~P.; Hofrichter,~J.; Easton,~W. \emph{Nature}
  \textbf{1997}, \emph{390}, 196--199\relax
\mciteBstWouldAddEndPuncttrue
\mciteSetBstMidEndSepPunct{\mcitedefaultmidpunct}
{\mcitedefaultendpunct}{\mcitedefaultseppunct}\relax
\EndOfBibitem
\bibitem[Hess \latin{et~al.}(2008)Hess, Kutzner, van~der Spoel, and
  Lindahl]{gromacs}
Hess,~B.; Kutzner,~C.; van~der Spoel,~D.; Lindahl,~E. GROMACS 4: Algorithms for
  highly efficient, load-balanced and scalable molecular simulation. \emph{J.
  Chem. Theory Comput.} \textbf{2008}, \emph{4}, 435--447\relax
\mciteBstWouldAddEndPuncttrue
\mciteSetBstMidEndSepPunct{\mcitedefaultmidpunct}
{\mcitedefaultendpunct}{\mcitedefaultseppunct}\relax
\EndOfBibitem
\bibitem[Laio and Parrinello(2002)Laio, and Parrinello]{metad}
Laio,~A.; Parrinello,~M. {Escaping free-energy minima}. \emph{{Proc. Natl.
  Acad. Sci. USA}} \textbf{2002}, \emph{99}, 12562--12566\relax
\mciteBstWouldAddEndPuncttrue
\mciteSetBstMidEndSepPunct{\mcitedefaultmidpunct}
{\mcitedefaultendpunct}{\mcitedefaultseppunct}\relax
\EndOfBibitem
\bibitem[Bonomi and Parrinello(2010)Bonomi, and Parrinello]{wte}
Bonomi,~M.; Parrinello,~M. Enhanced Sampling in the Well-Tempered Ensemble.
  \emph{Phys. Rev. Lett.} \textbf{2010}, \emph{104}, 190601\relax
\mciteBstWouldAddEndPuncttrue
\mciteSetBstMidEndSepPunct{\mcitedefaultmidpunct}
{\mcitedefaultendpunct}{\mcitedefaultseppunct}\relax
\EndOfBibitem
\bibitem[Bonomi \latin{et~al.}(2009)Bonomi, Branduardi, Bussi, Camilloni,
  Provasi, Raiteri, Donadio, Marinelli, Pietrucci, Broglia, and
  Parrinello]{plumed}
Bonomi,~M.; Branduardi,~D.; Bussi,~G.; Camilloni,~C.; Provasi,~D.; Raiteri,~P.;
  Donadio,~D.; Marinelli,~F.; Pietrucci,~F.; Broglia,~R.~A.; Parrinello,~M.
  PLUMED: A portable plugin for free-energy calculations with molecular
  dynamics. \emph{Comp. Phys. Comm.} \textbf{2009}, \emph{180}, 1961 --
  1972\relax
\mciteBstWouldAddEndPuncttrue
\mciteSetBstMidEndSepPunct{\mcitedefaultmidpunct}
{\mcitedefaultendpunct}{\mcitedefaultseppunct}\relax
\EndOfBibitem
\bibitem[Tribello \latin{et~al.}(2014)Tribello, Bonomi, Branduardi, Camilloni,
  and Bussi]{plumed2}
Tribello,~G.~A.; Bonomi,~M.; Branduardi,~D.; Camilloni,~C.; Bussi,~G.
  \{PLUMED\} 2: New feathers for an old bird. \emph{Comput. Phys. Commun.}
  \textbf{2014}, \emph{185}, 604 -- 613\relax
\mciteBstWouldAddEndPuncttrue
\mciteSetBstMidEndSepPunct{\mcitedefaultmidpunct}
{\mcitedefaultendpunct}{\mcitedefaultseppunct}\relax
\EndOfBibitem
\bibitem[Bonomi \latin{et~al.}(2009)Bonomi, Barducci, and
  Parrinello]{max-reweighting}
Bonomi,~M.; Barducci,~A.; Parrinello,~M. Reconstructing the equilibrium
  Boltzmann distribution from well-tempered metadynamics. \emph{J. Comput.
  Chem.} \textbf{2009}, \emph{30}, 1615--1621\relax
\mciteBstWouldAddEndPuncttrue
\mciteSetBstMidEndSepPunct{\mcitedefaultmidpunct}
{\mcitedefaultendpunct}{\mcitedefaultseppunct}\relax
\EndOfBibitem
\bibitem[Berteotti \latin{et~al.}(2011)Berteotti, Barducci, and
  Parrinello]{ale-urea}
Berteotti,~A.; Barducci,~A.; Parrinello,~M. Effect of Urea on the
  $\beta$-Hairpin Conformational Ensemble and Protein Denaturation Mechanism.
  \emph{J. Am. Chem. Soc.} \textbf{2011}, \emph{133}, 17200--17206\relax
\mciteBstWouldAddEndPuncttrue
\mciteSetBstMidEndSepPunct{\mcitedefaultmidpunct}
{\mcitedefaultendpunct}{\mcitedefaultseppunct}\relax
\EndOfBibitem
\bibitem[Zhou \latin{et~al.}({2001})Zhou, Berne, and Germain]{zhou-germain}
Zhou,~R.; Berne,~B.; Germain,~R. {The free energy landscape for beta hairpin
  folding in explicit water}. \emph{{Proc. Natl. Acad. Sci. U.S.A.}}
  \textbf{{2001}}, \emph{{98}}, {14931--14936}\relax
\mciteBstWouldAddEndPuncttrue
\mciteSetBstMidEndSepPunct{\mcitedefaultmidpunct}
{\mcitedefaultendpunct}{\mcitedefaultseppunct}\relax
\EndOfBibitem
\bibitem[Best and Mittal({2011})Best, and Mittal]{best-mittal}
Best,~R.~B.; Mittal,~J. {Microscopic events in beta-hairpin folding from
  alternative unfolded ensembles}. \emph{{Proc. Natl. Acad. Sci. U.S.A.}}
  \textbf{{2011}}, \emph{{108}}, {11087--11092}\relax
\mciteBstWouldAddEndPuncttrue
\mciteSetBstMidEndSepPunct{\mcitedefaultmidpunct}
{\mcitedefaultendpunct}{\mcitedefaultseppunct}\relax
\EndOfBibitem
\bibitem[Garcia and Sanbonmatsu({2001})Garcia, and Sanbonmatsu]{garc-sanb}
Garcia,~A.; Sanbonmatsu,~K. {Exploring the energy landscape of a beta hairpin
  in explicit solvent}. \emph{{Proteins: Struct., Funct., Genet.}}
  \textbf{{2001}}, \emph{{42}}, {345--354}\relax
\mciteBstWouldAddEndPuncttrue
\mciteSetBstMidEndSepPunct{\mcitedefaultmidpunct}
{\mcitedefaultendpunct}{\mcitedefaultseppunct}\relax
\EndOfBibitem
\bibitem[De~Sancho \latin{et~al.}(2013)De~Sancho, Mittal, and Best]{best}
De~Sancho,~D.; Mittal,~J.; Best,~R.~B. Folding Kinetics and Unfolded State
  Dynamics of the GB1 Hairpin from Molecular Simulation. \emph{J. Chem. Theory
  Comput.} \textbf{2013}, \emph{9}, 1743--1753\relax
\mciteBstWouldAddEndPuncttrue
\mciteSetBstMidEndSepPunct{\mcitedefaultmidpunct}
{\mcitedefaultendpunct}{\mcitedefaultseppunct}\relax
\EndOfBibitem
\bibitem[Bonomi \latin{et~al.}(2008)Bonomi, Branduardi, Gervasio, and
  Parrinello]{bonomi}
Bonomi,~M.; Branduardi,~D.; Gervasio,~F.~L.; Parrinello,~M. The Unfolded
  Ensemble and Folding Mechanism of the C-Terminal GB1 $\beta$-Hairpin.
  \emph{J. Am. Chem. Soc.} \textbf{2008}, \emph{130}, 13938--13944\relax
\mciteBstWouldAddEndPuncttrue
\mciteSetBstMidEndSepPunct{\mcitedefaultmidpunct}
{\mcitedefaultendpunct}{\mcitedefaultseppunct}\relax
\EndOfBibitem
\bibitem[Lindorff-Larsen \latin{et~al.}(2010)Lindorff-Larsen, Piana, Palmo,
  Maragakis, Klepeis, Dror, and Shaw]{amber-ildn}
Lindorff-Larsen,~K.; Piana,~S.; Palmo,~K.; Maragakis,~P.; Klepeis,~J.~L.;
  Dror,~R.~O.; Shaw,~D.~E. Improved side-chain torsion potentials for the Amber
  ff99SB protein force field. \emph{Proteins: Struct., Funct., Bioinf.}
  \textbf{2010}, \emph{78}, 1950--1958\relax
\mciteBstWouldAddEndPuncttrue
\mciteSetBstMidEndSepPunct{\mcitedefaultmidpunct}
{\mcitedefaultendpunct}{\mcitedefaultseppunct}\relax
\EndOfBibitem
\bibitem[Lindorff-Larsen \latin{et~al.}(2012)Lindorff-Larsen, Maragakis, Piana,
  Eastwood, Dror, and Shaw]{lindorff}
Lindorff-Larsen,~K.; Maragakis,~P.; Piana,~S.; Eastwood,~M.~P.; Dror,~R.~O.;
  Shaw,~D.~E. Systematic Validation of Protein Force Fields against
  Experimental Data. \emph{PLoS ONE} \textbf{2012}, \emph{7}, e32131\relax
\mciteBstWouldAddEndPuncttrue
\mciteSetBstMidEndSepPunct{\mcitedefaultmidpunct}
{\mcitedefaultendpunct}{\mcitedefaultseppunct}\relax
\EndOfBibitem
\bibitem[Rose \latin{et~al.}(2006)Rose, Fleming, Banavar, and
  Maritan]{backbone}
Rose,~G.~D.; Fleming,~P.~J.; Banavar,~J.~R.; Maritan,~A. \emph{Proc. Nat. Sci
  U.S.A.} \textbf{2006}, \emph{103}, 16623\relax
\mciteBstWouldAddEndPuncttrue
\mciteSetBstMidEndSepPunct{\mcitedefaultmidpunct}
{\mcitedefaultendpunct}{\mcitedefaultseppunct}\relax
\EndOfBibitem
\bibitem[Hegler \latin{et~al.}(2008)Hegler, Weinkam, and Wolynes]{spectrum}
Hegler,~J.~A.; Weinkam,~P.; Wolynes,~P.~G. \emph{HFSP Journal} \textbf{2008},
  \emph{2}, 307\relax
\mciteBstWouldAddEndPuncttrue
\mciteSetBstMidEndSepPunct{\mcitedefaultmidpunct}
{\mcitedefaultendpunct}{\mcitedefaultseppunct}\relax
\EndOfBibitem
\bibitem[Hegler \latin{et~al.}(2008)Hegler, Weinkam, and Wolynes]{funnel}
Hegler,~J.~A.; Weinkam,~P.; Wolynes,~P.~G. \emph{HFSP Journal} \textbf{2008},
  \emph{2}, 307\relax
\mciteBstWouldAddEndPuncttrue
\mciteSetBstMidEndSepPunct{\mcitedefaultmidpunct}
{\mcitedefaultendpunct}{\mcitedefaultseppunct}\relax
\EndOfBibitem
\bibitem[Noy \latin{et~al.}({2008})Noy, Kalisman, and Keasar]{noy-keasar}
Noy,~K.; Kalisman,~N.; Keasar,~C. {Prediction of structural stability of short
  beta-hairpin peptides by molecular dynamics and knowledge-based potentials}.
  \emph{{BMC Struct. Biol.}} \textbf{{2008}}, \emph{{8}}\relax
\mciteBstWouldAddEndPuncttrue
\mciteSetBstMidEndSepPunct{\mcitedefaultmidpunct}
{\mcitedefaultendpunct}{\mcitedefaultseppunct}\relax
\EndOfBibitem
\bibitem[Juraszek and Bolhuis({2009})Juraszek, and Bolhuis]{jura-bolh}
Juraszek,~J.; Bolhuis,~P.~G. {Effects of a Mutation on the Folding Mechanism of
  beta-Hairpin}. \emph{{J. Phys. Chem. B}} \textbf{{2009}}, \emph{{113}},
  {16184--16196}\relax
\mciteBstWouldAddEndPuncttrue
\mciteSetBstMidEndSepPunct{\mcitedefaultmidpunct}
{\mcitedefaultendpunct}{\mcitedefaultseppunct}\relax
\EndOfBibitem
\bibitem[F.Hofmeister(1888)]{hoffmeister}
F.Hofmeister, \emph{Arch. Exp. Pathol. Pharmacol.} \textbf{1888}, \emph{24},
  247--260\relax
\mciteBstWouldAddEndPuncttrue
\mciteSetBstMidEndSepPunct{\mcitedefaultmidpunct}
{\mcitedefaultendpunct}{\mcitedefaultseppunct}\relax
\EndOfBibitem
\bibitem[Oliverberg and Wolynes(2008)Oliverberg, and Wolynes]{wolynes}
Oliverberg,~M.; Wolynes,~P.~G. \emph{Quaterly Reviews of Biophyics}
  \textbf{2008}, \emph{71}, 245\relax
\mciteBstWouldAddEndPuncttrue
\mciteSetBstMidEndSepPunct{\mcitedefaultmidpunct}
{\mcitedefaultendpunct}{\mcitedefaultseppunct}\relax
\EndOfBibitem
\end{mcitethebibliography}
\end{document}